
%
%


\magnification=\magstephalf
\hsize=13.0 true cm
\vsize=19 true cm
\hoffset=1.50 true cm
\voffset=2.0 true cm

\abovedisplayskip=12pt plus 3pt minus 3pt
\belowdisplayskip=12pt plus 3pt minus 3pt
\parindent=2em


\font\sixrm=cmr6
\font\eightrm=cmr8
\font\ninerm=cmr9

\font\sixi=cmmi6
\font\eighti=cmmi8
\font\ninei=cmmi9

\font\sixsy=cmsy6
\font\eightsy=cmsy8
\font\ninesy=cmsy9

\font\sixbf=cmbx6
\font\eightbf=cmbx8
\font\ninebf=cmbx9

\font\eightit=cmti8
\font\nineit=cmti9

\font\eightsl=cmsl8
\font\ninesl=cmsl9

\font\sixss=cmss8 at 8 true pt
\font\sevenss=cmss9 at 9 true pt
\font\eightss=cmss8
\font\niness=cmss9
\font\tenss=cmss10

\font\bigrm=cmr10 at 12 true pt
\font\bigbf=cmbx10 at 12 true pt

\catcode`@=11
\newfam\ssfam

\def\tenpoint{\def\rm{\fam0\tenrm}%
    \textfont0=\tenrm \scriptfont0=\sevenrm \scriptscriptfont0=\fiverm
    \textfont1=\teni  \scriptfont1=\seveni  \scriptscriptfont1=\fivei
    \textfont2=\tensy \scriptfont2=\sevensy \scriptscriptfont2=\fivesy
    \textfont3=\tenex \scriptfont3=\tenex   \scriptscriptfont3=\tenex
    \textfont\itfam=\tenit                  \def\it{\fam\itfam\tenit}%
    \textfont\slfam=\tensl                  \def\sl{\fam\slfam\tensl}%
    \textfont\bffam=\tenbf \scriptfont\bffam=\sevenbf
    \scriptscriptfont\bffam=\fivebf
                                            \def\bf{\fam\bffam\tenbf}%
    \textfont\ssfam=\tenss \scriptfont\ssfam=\sevenss
    \scriptscriptfont\ssfam=\sevenss
                                            \def\ss{\fam\ssfam\tenss}%
    \normalbaselineskip=13pt
    \setbox\strutbox=\hbox{\vrule height8.5pt depth3.5pt width0pt}%
    \let\big=\tenbig
    \normalbaselines\rm}

\def\ninepoint{\def\rm{\fam0\ninerm}%
    \textfont0=\ninerm      \scriptfont0=\sixrm
                            \scriptscriptfont0=\fiverm
    \textfont1=\ninei       \scriptfont1=\sixi
                            \scriptscriptfont1=\fivei
    \textfont2=\ninesy      \scriptfont2=\sixsy
                            \scriptscriptfont2=\fivesy
    \textfont3=\tenex       \scriptfont3=\tenex
                            \scriptscriptfont3=\tenex
    \textfont\itfam=\nineit \def\it{\fam\itfam\nineit}%
    \textfont\slfam=\ninesl \def\sl{\fam\slfam\ninesl}%
    \textfont\bffam=\ninebf \scriptfont\bffam=\sixbf
                            \scriptscriptfont\bffam=\fivebf
                            \def\bf{\fam\bffam\ninebf}%
    \textfont\ssfam=\niness \scriptfont\ssfam=\sixss
                            \scriptscriptfont\ssfam=\sixss
                            \def\ss{\fam\ssfam\niness}%
    \normalbaselineskip=12pt
    \setbox\strutbox=\hbox{\vrule height8.0pt depth3.0pt width0pt}%
    \let\big=\ninebig
    \normalbaselines\rm}

\def\eightpoint{\def\rm{\fam0\eightrm}%
    \textfont0=\eightrm      \scriptfont0=\sixrm
                             \scriptscriptfont0=\fiverm
    \textfont1=\eighti       \scriptfont1=\sixi
                             \scriptscriptfont1=\fivei
    \textfont2=\eightsy      \scriptfont2=\sixsy
                             \scriptscriptfont2=\fivesy
    \textfont3=\tenex        \scriptfont3=\tenex
                             \scriptscriptfont3=\tenex
    \textfont\itfam=\eightit \def\it{\fam\itfam\eightit}%
    \textfont\slfam=\eightsl \def\sl{\fam\slfam\eightsl}%
    \textfont\bffam=\eightbf \scriptfont\bffam=\sixbf
                             \scriptscriptfont\bffam=\fivebf
                             \def\bf{\fam\bffam\eightbf}%
    \textfont\ssfam=\eightss \scriptfont\ssfam=\sixss
                             \scriptscriptfont\ssfam=\sixss
                             \def\ss{\fam\ssfam\eightss}%
    \normalbaselineskip=10pt
    \setbox\strutbox=\hbox{\vrule height7.0pt depth2.0pt width0pt}%
    \let\big=\eightbig
    \normalbaselines\rm}

\def\tenbig#1{{\hbox{$\left#1\vbox to8.5pt{}\right.\n@space$}}}
\def\ninebig#1{{\hbox{$\textfont0=\tenrm\textfont2=\tensy
                       \left#1\vbox to7.25pt{}\right.\n@space$}}}
\def\eightbig#1{{\hbox{$\textfont0=\ninerm\textfont2=\ninesy
                       \left#1\vbox to6.5pt{}\right.\n@space$}}}

\font\sectionfont=cmbx10
\font\subsectionfont=cmti10

\def\figurecaptionfont{\ninepoint}
\def\tablecaptionfont{\ninepoint}
\def\footnotefont{\eightpoint}


\newcount\equationno
\newcount\bibitemno
\newcount\figureno
\newcount\tableno

\equationno=0
\bibitemno=0
\figureno=0
\tableno=0
\advance\pageno by -1


\footline={\ifnum\pageno=0{\hfil}\else
{\hss\rm\the\pageno\hss}\fi}


\def\section #1. #2 \par
{\vskip0pt plus .20\vsize\penalty-150 \vskip0pt plus-.20\vsize
\vskip 1.6 true cm plus 0.2 true cm minus 0.2 true cm
\global\def\equationlabel{#1}
\global\equationno=0
\centerline{\sectionfont #1. #2}\par
\immediate\write\terminal{Section #1. #2}
\vskip 0.7 true cm plus 0.1 true cm minus 0.1 true cm}


\def\subsection #1 \par
{\vskip0pt plus .15\vsize\penalty-50 \vskip0pt plus-.15\vsize
\vskip2.5ex plus 0.1ex minus 0.1ex
\leftline{\subsectionfont #1}\par
\immediate\write\terminal{Subsection #1}
\vskip1.0ex plus 0.1ex minus 0.1ex
\noindent}


\def\appendix #1 \par
{\vskip0pt plus .20\vsize\penalty-150 \vskip0pt plus-.20\vsize
\vskip 1.6 true cm plus 0.2 true cm minus 0.2 true cm
\global\def\equationlabel{\hbox{\rm#1}}
\global\equationno=0
\centerline{\sectionfont Appendix #1}\par
\immediate\write\terminal{Appendix #1}
\vskip 0.7 true cm plus 0.1 true cm minus 0.1 true cm}


\def\enum{\global\advance\equationno by 1
(\equationlabel.\the\equationno)}


\def\ifundefined#1{\expandafter\ifx\csname#1\endcsname\relax}

\def\ref#1{\ifundefined{#1}?\immediate\write\terminal{unknown reference
on page \the\pageno}\else\csname#1\endcsname\fi}

\newwrite\terminal
\newwrite\bibitemlist

\def\bibitem#1#2\par{\global\advance\bibitemno by 1
\immediate\write\bibitemlist{\string\def
\expandafter\string\csname#1\endcsname
{\the\bibitemno}}
\item{[\the\bibitemno]}#2\par}

\def\beginbibliography{
\vskip0pt plus .20\vsize\penalty-150 \vskip0pt plus-.20\vsize
\vskip 1.6 true cm plus 0.2 true cm minus 0.2 true cm
\centerline{\sectionfont References}\par
\immediate\write\terminal{References}
\immediate\openout\bibitemlist=biblist
\frenchspacing
\vskip 0.7 true cm plus 0.1 true cm minus 0.1 true cm}

\def\endbibliography{
\immediate\closeout\bibitemlist
\nonfrenchspacing}


\def\figurecaption#1{\global\advance\figureno by 1
\narrower\figurecaptionfont
Fig.~\the\figureno. #1}

\def\tablecaption#1{\global\advance\tableno by 1
\vbox to 0.5 true cm { }
\centerline{\tablecaptionfont%
Table~\the\tableno. #1}
\vskip-0.4 true cm}

\tenpoint


\def\blackboardrrm{\mathchoice
{\rm I\kern-0.21 em{R}}{\rm I\kern-0.21 em{R}}
{\rm I\kern-0.19 em{R}}{\rm I\kern-0.19 em{R}}}

\def\blackboardzrm{\mathchoice
{\rm Z\kern-0.32 em{Z}}{\rm Z\kern-0.32 em{Z}}
{\rm Z\kern-0.28 em{Z}}{\rm Z\kern-0.28 em{Z}}}

\def\blackboardh{\mathchoice
{\ss I\kern-0.14 em{H}}{\ss I\kern-0.14 em{H}}
{\ss I\kern-0.11 em{H}}{\ss I\kern-0.11 em{H}}}

\def\blackboardp{\mathchoice
{\ss I\kern-0.14 em{P}}{\ss I\kern-0.14 em{P}}
{\ss I\kern-0.11 em{P}}{\ss I\kern-0.11 em{P}}}

\def\blackboardt{\mathchoice
{\ss T\kern-0.52 em{T}}{\ss T\kern-0.52 em{T}}
{\ss T\kern-0.40 em{T}}{\ss T\kern-0.40 em{T}}}

\def\thicktablerule{\hrule height1pt}
\def\thintablerule{\hrule height0.4pt}

\input biblist.tex



\def\rmd{{\rm d}}
\def\rmD{{\rm D}}
\def\rme{{\rm e}}
\def\rmO{{\rm O}}



\def\proof{\noindent{\sl Proof:}\kern0.6em}

\def\frac#1#2{\hbox{$#1\over#2$}}
\def\dual{\mathstrut^*\kern-0.1em}


\def\MeV{{\rm MeV}}
\def\GeV{{\rm GeV}}

\def\fm{{\rm fm}}



\def\gbar{\bar{g}}

\def\gtilde{\tilde{g}_0}

\def\msbar{{\rm \overline{MS\kern-0.14em}\kern0.14em}}

\def\qqbar{{\rm q\bar{q}}}


\def\SUtwo{{\rm SU}(2)}
\def\SUthree{{\rm SU}(3)}

\def\tr{{\rm tr}}


\def\schrodinger{{\cal Z}}
\def\effaction{\Gamma}


\def\bvalue{C}
\def\bfield{B}
\def\bfieldparm{\eta}
\def\bvaluelat{W}
\def\bfieldlat{V}


\def\stringtension{K}


\def\Nor{N}
\def\tauint{\tau}

%
\line{\hfill CERN-TH 6566/92}
\line{\hfill DESY-92-096}
\vskip 1 true cm
\centerline
{\bigbf Computation of the Running Coupling in the}
\vskip 0.2 true cm
\centerline
{\bigbf SU(2) Yang-Mills Theory}
\vskip 1.3 true cm
\centerline{\bigrm Martin L\"{u}scher}
\vskip 2ex
\centerline{Deutsches Elektronen-Synchrotron DESY}
\centerline{Notkestrasse 85, D-2000 Hamburg 52, Germany}
\vskip 0.7 true cm
\centerline{\bigrm Rainer Sommer and Ulli Wolff}
\vskip 2ex
\centerline{CERN, Theory Division, CH-1211 Gen\`eve 23, Switzerland}
\vskip 0.7 true cm
\centerline{\bigrm Peter Weisz}
\vskip 2ex
\centerline{Max Planck Institut f\"ur Physik}
\centerline{F\"ohringerring 6, D-8000 M\"unchen 40, Germany}
\vskip 1.3 true cm
\centerline{\bf Abstract}
\vskip 1.5ex
A finite-size scaling technique is applied to
the $\SUtwo$ gauge theory
(without matter fields)
to compute a non-perturbatively defined running coupling
$\alpha(q)$ for a range of momenta $q$ given in units
of the string tension $\stringtension$.
We find that already at rather low $q$, the evolution of
$\alpha(q)$ is well described by the 2-loop approximation
to the Callan-Symanzik $\beta$--function.
At the highest momentum reached,
$q=20\times\sqrt{\stringtension}$,
we obtain
$\alpha_{\msbar}(q)=0.187\pm0.005\pm0.009$
for the running coupling in the $\msbar$ scheme of
dimensional regularization.
\vskip 1 true cm
\line{CERN-TH 6566/92 \hfill}
\line{DESY-92-096 \hfill}
\line{July 1992 \hfill}
\vfill
\eject

\section 1. Introduction

{}From the point of view of perturbation theory,
the renormalized coupling $\alpha(q)$ in QCD is an input
parameter, whose value at some reference momentum $q=q_0$
must be supplied by experiment.
There is little doubt that QCD is a well-defined theory
also at low energies, where the perturbation
expansion does not apply.
We may thus imagine that the parameters of the theory are fixed
through the
hadron spectrum, for example, or some other set of
experimentally accessible quantities in the low-energy domain.
The running coupling then becomes a
computable function of momentum, in any
renormalization scheme that one may choose.

A theoretical determination of $\alpha(q)$ at high
energies is obviously desirable. In particular,
one would be interested to know at which scale the perturbative
evolution of the coupling sets in.
Since one is concerned with the non-perturbative properties of QCD,
the lattice formulation of the theory,
combined with numerical simulation techniques,
is currently the most promising way to approach the problem
[\ref{Wilson}--\ref{MiGbar}].
The basic difficulty in any such calculation is that
the momenta $q$ of interest can be orders of magnitude greater than
the masses of the light particles in the theory.
Lattices sufficiently wide to avoid finite volume effects
and with a spacing $a$ substantially
smaller than $1/q$ thus tend to be much larger than what
can be simulated on a computer.

Many years ago Wilson pointed out that
this difficulty may be overcome, in principle,
by introducing a renormalization group transformation
which allows one to step up
the energy scale in a recursive manner
[\ref{Wilson}].
No simulation of an exceedingly large lattice
would then be required, while
all physical scales are kept at
a safe distance from the ultra-violet cutoff $1/a$.

The finite-size scaling technique described in
ref.[\ref{LueWeWo}] may be regarded as a particular realization
of this basic idea, even though the details are
quite different from Wilson's formulation.
The method has already been shown to work well in the
case of the two-dimensional non-linear $\sigma$--model.
In the present paper it is applied
to the pure $\SUtwo$ gauge theory, the simplest asymptotically
free theory in four dimensions.
As a result we shall be able to compute
the running coupling (in a certain adapted scheme [\ref{LueNaWeWo}])
over a large range of momenta $q$,
reaching energies far above the scale set by the string tension.

Other strategies to compute the running coupling in
non-Abelian gauge theories have recently been put forward
by El-Khadra et al.~[\ref{AidaEtAl}]
and Michael [\ref{MiGbar}].
In both cases one considers a single lattice which covers
all relevant distance scales, and one is, therefore,
limited to rather low momenta $q$.
In ref.[\ref{MiGbar}], for example, the coupling is determined
from the static quark potential at short distances.
The largest lattice currently available has $56\times 48^3$
points and a spacing $a$ roughly equal to
$0.03\,\fm$ [\ref{UKQCDI}].
The distances at which the coupling
can be calculated with some confidence are thus greater than
$0.1\,\fm$ or so, a limitation which will be difficult to alleviate.


\section 2. Finite-Size Technology

The finite-size scaling technique of ref.[\ref{LueWeWo}] is
based on a renormalized coupling
$$
  \alpha(q)={\gbar^2(L)\over4\pi},\qquad q=1/L,
  \eqno\enum
$$
which runs
with the linear extent $L$ of the lattice.
As proposed in ref.[\ref{LueNaWeWo}],
we define $\gbar^2(L)$
through the response of the system
to a constant colour-electric background field.
For detailed explanations the reader is referred to
refs.[\ref{LueWeWo},\ref{LueNaWeWo}].
Here we only list the basic definitions
and outline our strategy.

\subsection 2.1 Boundary conditions and functional integral

We choose to set up the theory on a hyper-cubic euclidean
lattice with spacing $a$ and size $L\times L\times L\times L$.
In particular, the possible values of the time coordinate $x^0$
of a lattice point $x$ are $x^0=0,a,2a,\ldots,L$
($L$ is taken to be an integer multiple of $a$).
The spatial sublattices at fixed times are thought to be
wrapped on a torus, i.e.~we assume
periodic boundary conditions in these directions.

A gauge field $U$ on the lattice is an assignment of
a matrix $U(x,\mu)\in\SUtwo$
to every pair $(x,x+a\hat{\mu})$ of nearest neighbor lattice points
($\hat{\mu}$ denotes the unit vector in the $\mu$--direction
and $\mu=0,1,2,3$).
At the top and bottom of the lattice, the link variables
are required to satisfy inhomogenous Dirichlet boundary conditions,
$$
  \left.U(x,k)\right|_{x^0=0}=
  \bvaluelat({\bf x},k),
  \qquad
  \left.U(x,k)\right|_{x^0=L}=
  \bvaluelat'({\bf x},k),
  \eqno\enum
$$
for all $k=1,2,3$,
where $\bvaluelat$ and $\bvaluelat'$ are
prescribed spatial gauge fields.
They will be set to some particular values below.

The action of a lattice gauge field
$U$ is taken to be
$$
  S[U]=
  {1\over g_0^2}\sum_p w(p)\,
  \tr\left\{1-U(p)\right\},
  \eqno\enum
$$
with $g_0$ being the bare coupling.
The sum in eq.(2.3) runs over all
{\it oriented} plaquettes $p$
on the lattice and $U(p)$ denotes the parallel transporter
around $p$.
The weight $w(p)$ is equal to 1 in all cases
except for the spatial plaquettes
at $x^0=0$ and $x^0=L$
which are given the weight ${1\over2}$.

The partition function of the system,
$$
  \schrodinger=
  \int\rmD[U]\,\rme^{-S[U]},
  \qquad \rmD[U]=\prod_{x,\mu}\rmd U(x,\mu),
  \eqno\enum
$$
involves an integration
over all fields $U$ with fixed boundary values
$\bvaluelat$ and $\bvaluelat'$.
$\schrodinger$ is also referred to
as the Schr\"odinger functional,
because it is just
the (euclidean) propagation kernel for
going from the initial field configuration $\bvaluelat$
at time $x^0=0$ to
the final configuration $\bvaluelat'$ at $x^0=L$.

\subsection 2.2 Background field

At small couplings $g_0$, the integral (2.4)
is dominated by the field configurations with least action.
In the cases considered below,
there is only one such configuration
(modulo gauge transformations).
Through the boundary conditions, we have thus forced
a background field into the system.

As indicated at the beginning of this section,
we are interested in generating a
constant colour-electric background field.
This can be achieved by choosing the boundary values
$\bvaluelat$ and $\bvaluelat'$ to be
constant diagonal matrices. More precisely, we set
$$
  \eqalignno{
  \bvaluelat({\bf x},k)
  &=\exp\left\{a\bvalue^{\vphantom{\prime}}_k\right\},
  \qquad
  \bvalue^{\vphantom{\prime}}_k
  =\bfieldparm\,
  {\tau_3\over iL},
  &\enum\cr
  \noalign{\vskip1ex}
  \bvaluelat'({\bf x},k)
  &=\exp\left\{a\bvalue'_k\right\},
  \qquad
  \bvalue'_k
  =(\pi-\bfieldparm)\,
  {\tau_3\over iL},
  &\enum\cr}
$$
where $\bfieldparm$ is a real parameter, to be fixed later,
and $\tau_3$ the third Pauli matrix.
The induced background field is then given by
$$
  \bfieldlat(x,\mu)
  =\exp\left\{a\bfield_{\mu}(x)\right\},
  \eqno\enum
$$
with
$$
  \bfield_{0}
  =0,\qquad
  \bfield_k=\left[x^0\bvalue'_k
  +(L-x^0)\bvalue^{\vphantom{\prime}}_k\right]/L.
  \eqno\enum
$$
As shown
in ref.[\ref{LueNaWeWo}], this is indeed
a configuration with least action for the specified
boundary values, provided
$$
  0<\bfieldparm<\pi
  \quad\hbox{and}\quad
  L/a\geq 4.
  \eqno\enum
$$
{}From the linear time dependence of $\bfield$, it is
evident that $\bfieldlat$
corresponds to
a constant colour-electric
field.

\subsection 2.3 Definition of $\gbar^2(L)$

The effective action of the background field (2.7) is defined by
$$
  \effaction=-\ln\schrodinger.
  \eqno\enum
$$
To leading order in perturbation theory, $\effaction$ is simply equal
to the classical action
$$
  S[\bfieldlat]=
  {6\over g_0^2}
  \left\{{2L^2\over a^2}
  \sin\left[{a^2\over2L^2}(\pi-2\bfieldparm)\right]\right\}^2,
  \eqno\enum
$$
and the higher order corrections may be worked out
by expanding the Schr\"o\-dinger functional about the background
field.

A crucial observation now is
that the continuum limit of the effective action exists,
provided the bare coupling is
renormalized in the usual way.
As discussed in ref.[\ref{LueNaWeWo}],
this follows from power counting and gauge invariance.
The statement has, furthermore, been checked explicitly
to one-loop order.

We are thus led to define
a renormalized coupling $\gbar^2(L)$ through
$$
  {\partial\effaction\over\partial\bfieldparm}
  ={k\over\gbar^2(L)},
  \eqno\enum
$$
where the proportionality constant $k$ is adjusted such that
$\gbar^2(L)$ is equal to $g_0^2$ to lowest order.
{}From eq.(2.11) we deduce
$$
  k=-24\,{L^2\over a^2}
  \sin\left[{a^2\over L^2}(\pi-2\bfieldparm)\right].
  \eqno\enum
$$
A derivative with respect to $\bfieldparm$ is taken,
because expectation values are
much easier to compute
than partition functions
(cf.~sect.~3).

To complete the definition of $\gbar^2(L)$
we must finally pick some value for the background field
parameter $\bfieldparm$.
We decided to take
$$
  \bfieldparm=\pi/4
  \eqno\enum
$$
which is half-way between the zero action point $\bfieldparm=\pi/2$
and the boundary of the stability interval (2.9).

\subsection 2.4 Relation to other schemes

In the continuum limit, and at sufficiently high energies (small $L$),
perturbation theory may be used to relate different running couplings.
In particular, the connection between
the $\msbar$ scheme of dimensional regularization [\ref{BaBuDuMu}]
and our finite volume scheme is
[\ref{LueNaWeWo}]
$$
  \alpha_{\msbar}
  =\alpha+k_1\alpha^2+\ldots,
  \qquad
  k_1
  =0.94327(5)
  \eqno\enum
$$
[cf.~eq.(2.1); both couplings are at the same momentum $q$\/].

Another coupling in infinite volume,
$\alpha_{\qqbar}(q)$,
is defined by
$$
  \alpha_{\qqbar}(q)=\frac{4}{3}r^2F(r),
  \qquad
  q=1/r,
  \eqno\enum
$$
where $F(r)$ denotes
the force between static quarks at distance $r$.
Combining the expansion above with the one-loop results of
refs.[\ref{Fi},\ref{Bi}],
one finds
$$
  \alpha_{\qqbar}
  =\alpha+h_1\alpha^2+\ldots,
  \qquad
  h_1
  =0.99802(5),
  \eqno\enum
$$
i.e.~to this order, there is practically no difference
between $\alpha_{\qqbar}$ and $\alpha_{\msbar}$.

\subsection 2.5 Low-energy regime

When $L$ is greater than the
confinement scale,
the behaviour of the running coupling is determined
by non-perturbative effects.
An important observation at this point is that
the boundary fields
$\bvaluelat$ and $\bvaluelat'$ are locally pure gauge
configurations.
Any dependence of the effective action $\effaction$
on the background field parameter
$\bfieldparm$ is hence associated
with correlations ``around the world".
Since these are exponentially suppressed in a massive theory,
we conclude that
$\gbar^2(L)\propto\rme^{\mu L}$ at large $L$.

The mass
$\mu$ occurring here is characteristic for the dynamics of
the gauge fields
close to the boundaries of the lattice.
In particular, its relation to the bulk correlation length
is not obvious.
To make contact with the physical scales,
some extra work will therefore be needed
(cf.~sect.~4).

As a reference energy scale in infinite volume, we shall
take the string tension $\stringtension$
which is defined by
$$
  \stringtension=\lim_{r\to\infty}F(r).
  \eqno\enum
$$
There is no fundamental reason for choosing this particular
quantity.
Alternatives would be the energy splitting between
the 1P and 1S (quenched) charmonium levels
[\ref{AidaEtAl}], or the distance $r$ at which
$r^2F(r)=5$, for example.
[The force between physical heavy quarks
is known to be approximately equal to
$1\,\GeV/\fm$ at $r=1\,\fm$ so that $r^2F(r)=5$
at this point.]

\subsection 2.6 Renormalization group $\dag$

We again assume that the continuum limit has been taken
and define
the Callan-Symanzik $\beta$--function through
\footnote{$\phantom{\dag}$}{\footnotefont\kern-1em $\dag$
Our notation here is slightly different from the one
employed in ref.[\ref{LueWeWo}].
The reason for this is that the conventional normalization
of the $\Lambda$--parameter is unfortunately not the same
in gauge theories and two-dimensional non-linear $\sigma$--models.}
$$
  \beta(\gbar)=
  -L{\partial \gbar\over\partial L}.
  \eqno\enum
$$
{}From eq.(2.15) and the known perturbation expansion
of the $\beta$--function in the $\msbar$ scheme, we infer that
$$
  \beta(g)
  \mathrel{\mathop\sim_{g\to0}}
  -g^3\sum_{n=0}^{\infty}b_n g^{2n}
  \eqno\enum
$$
with
[\ref{GroWi}--\ref{Cas}]
$$
  b_0=\frac{11}{3}\left(8\pi^2\right)^{-1},\qquad
  b_1=\frac{34}{3}\left(8\pi^2\right)^{-2}.
  \eqno\enum
$$
The 3-loop coefficient $b_2$ depends on our choice of running coupling
and is presently not available.

When integrated towards short distances,
eq.(2.19) yields the asymptotic expression
$$
  \gbar^2(L)=
  {1\over b_0t}-{b_1\ln t\over b_0^3 t^2}+\rmO(t^{-3}(\ln t)^2),
  \qquad
  t=-\ln(\Lambda L)^2,
  \eqno\enum
$$
where $\Lambda$ is an integration constant, the $\Lambda$--parameter.

In the following a key r\^ole is
played by the step scaling function $\sigma(s,u)$.
For any given scale factor $s$ and initial value
$u=\gbar^2(L)$, the coupling
$u'=\gbar^2(sL)$ may be computed by integrating
the renormalization group equation (2.19)
(assuming the $\beta$--function is known).
$u'$ is a well-determined function of
$s$ and $u$, and so we may define
$$
  \sigma(s,u)=u'.
  \eqno\enum
$$
In other words, the step scaling function is an integrated
form of the $\beta$--function, which tells us what happens to
the coupling if the box size is changed by a factor $s$.

It is possible to calculate the step scaling function
through numerical simulation of the lattice theory.
To this end one chooses some value for the bare coupling
$g_0^2$ and
simulates two lattices with size $L$ and
$L'=sL$.
The coupling $u'=\gbar^2(L')$ then
provides an approximation to
$\sigma(s,u)$ at $u=\gbar^2(L)$.

To understand how good the approximation is, we first note
that for all lattice spacings a functional dependence
$$
  u'=\Sigma(s,u,a/L)
  \eqno\enum
$$
exists, which is obtained by eliminating $g_0^2$ in favour
of $u$.
{}From the discussion of the cutoff dependence
of the effective action in ref.[\ref{LueNaWeWo}],
we expect that $\Sigma(s,u,a/L)$ converges to
$\sigma(s,u)$ in the continuum limit $a\to0$
with a rate roughly proportional to $a/L$.

\topinsert
\newdimen\digitwidth
\setbox0=\hbox{\rm 0}
\digitwidth=\wd0
\catcode`@=\active
\def@{\kern\digitwidth}
\tablecaption{
One-loop coefficient
$\delta_1(a/L)$ in the expansion (2.25)
}
\vskip1ex
$$\vbox{\settabs\+  xx$10$
                 &  xxx$1.00000$
                 &  xxxxxx$10$
                 &  xxx$1.00000$
                 &  xx\cr
\vskip0.5ex
\thicktablerule
\vskip1ex
                \+  \hfill $L/a$              \hskip0.0ex
                 &  \hfill $\delta_1$         \hskip2.5ex
                 &  \hfill $L/a$              \hskip0.0ex
                 &  \hfill $\delta_1$         \hskip2.5ex
                 &  \quad\cr
\vskip 0.5ex
\thintablerule
\vskip 1ex
  \+ \hfill$ 6$ & \hfill$0.00623$ &
     \hfill$12$ & \hfill$0.00396$ & \cr
  \+ \hfill$ 8$ & \hfill$0.00540$ &
     \hfill$14$ & \hfill$0.00347$ & \cr
  \+ \hfill$10$ & \hfill$0.00459$ &
     \hfill$16$ & \hfill$0.00309$ & \cr
\vskip 1ex
\thicktablerule}$$
\endinsert

A more quantitative impression on the size of the cutoff effects
may be obtained in perturbation theory.
In particular, the relative deviation
$$
  \delta(u,a/L)={\Sigma(2,u,a/L)-\sigma(2,u)\over\sigma(2,u)}
  =\delta_1(a/L)\,u+\delta_2(a/L)\,u^2+\ldots,
  \eqno\enum
$$
turns out to be quite small at one-loop order (see table 1).
It will nevertheless be necessary to extrapolate
the numerical data to the continuum limit by simulating
a sequence of lattice pairs with decreasing lattice
spacings (and fixed coupling $u$).

\subsection 2.7 Strategy

Our principal aim is to compute $\gbar^2(L)$ for a range of
box sizes $L$ connecting the
low-energy domain with the perturbative scaling region.
This is achieved by setting up a recursion
$$
  u_{i+1}=\sigma(s_i,u_i),\qquad i=0,1,2,\ldots,
  \eqno\enum
$$
starting from some initial value $u_0$ of the
running coupling.
By construction we have
$$
  u_i=\gbar^2(L_i),
  \qquad
  L_i=L_0\prod_{j=0}^{i-1}s_j,
  \eqno\enum
$$
for some box size $L_0$.
Our choice of initial value $u_0$ and scale factors $s_i$
will be such that the recursion progresses from the perturbative
small coupling regime towards larger box sizes.

Of course the statistical and extrapolation errors limit
the number of iterations that can be done in practice.
A careful discussion of the error propagation is certainly
necessary, and we shall come back to this issue when we
analyse our data in sect.~4.

After a certain number $n$
of iterations, depending on the initial
value $u_0$, the final box size $L_n$ will be
close to or even larger than the scale set by
the string tension
$\stringtension$.
At this point it is possible to determine
the dimensionless combination $L_n\sqrt{\stringtension}$.
As a result one has calculated the running coupling
at all $L_i$ given in units of the string tension.

At the lower end of the range of $L$ covered,
where the coupling is small,
we may finally apply perturbation theory
to determine the value of any other coupling such as
$\alpha_{\msbar}$ or $\alpha_{\qqbar}$
(cf.~subsect.~2.4).
Note that all reference to a finite volume drops out in this last step.
One simply
gets $\alpha_{\msbar}$ at some large momenta $q_i$ given in units of
the string tension.


\section 3. Numerical Simulation

For fixed boundary values $\bvaluelat$ and $\bvaluelat'$,
the system defined through the partition function (2.4) can be
simulated by adapting any one of the known Monte Carlo algorithms
for pure gauge theories on periodic lattices.
The simulations that we have performed are here described in
some detail and a complete list of our data
on the running coupling is given.
These will be further analysed in sect.~4.

\subsection 3.1 Observables

{}From the definition (2.12) of the running coupling
it follows that
$$
  \gbar^2(L)=k
  \left\langle\partial S/\partial\bfieldparm\right\rangle^{-1}.
  \eqno\enum
$$
The field variables integrated over do not depend on
the background field parameter $\bfieldparm$
and so only the plaquettes touching the boundary contribute to
$\partial S/\partial\bfieldparm$.
The observable we
shall be concerned with in the following is thus given by
$$
  {\partial S\over\partial\bfieldparm}=
  -{2a^3\over g_0^2 L}\sum_{\bf x}\sum_{l=1}^3
  \left\{E'_l({\bf x})+E^{\vphantom{\prime}}_l({\bf x})\right\},
  \eqno\enum
$$
where $E'_l$ and $E^{\vphantom{\prime}}_l$ denote the
$\tau_3$-component of the colour-electric
field at the top and bottom of the lattice.
In particular,
$$
  E_l({\bf x})=
  {1\over i a^2} \tr\left\{
  \tau_3
  \bvaluelat({\bf x},l)
  U(x+a\hat{l},0)
  U(x+a\hat{0},l)^{-1}
  U(x,0)^{-1}
  \right\}_{x^0=0},
  \eqno\enum
$$
and a similar expression is obtained for $E'_l$.

It is conceivable that other observables exist
which have the same expectation value as
$\partial S/\partial\bfieldparm$ but a
significantly smaller variance.
We were, however, not successful in our search for
such an ``improved" observable.
In particular,
the so-called multi-hit method [\ref{hit}], when applied to the
dynamical link variables in eq.(3.3),
did not result in any appreciable increase in efficiency.
The data listed below have thus been generated taking
$\partial S/\partial\bfieldparm$ as the observable.

\topinsert
\vbox{\vskip 10 true cm plus 0.2 true cm}
\figurecaption{
Integrated autocorrelation times
$\tauint$ for the observable
$\partial S/\partial\bfieldparm$,
given in numbers of update sweeps through the lattice
(counting over-relaxation and heatbath sweeps).}
\endinsert

\subsection 3.2 Monte Carlo algorithm

The most efficient simulation algorithms
for pure gauge theories
known today involve the idea of over-relaxation
in one form or the other
[\ref{ORI}--\ref{ORIII}].
Our algorithm is a hybrid one, with
$\Nor$ exactly microcanonical sweeps through the lattice
followed by 1 heatbath update pass.
Geometrically the program is organized
in time slices which are visited sequentially.
At any given time, the lattice is divided into
2 or 4 sublattices, depending on whether $L/a$ is even or odd.
This is done in such a way that the link variables
in a fixed
sublattice are decoupled and so can be updated
in a vector mode.


\topinsert
%
\newdimen\digitwidth
\setbox0=\hbox{\rm 0}
\digitwidth=\wd0
\catcode`@=\active
\def@{\kern\digitwidth}
\tablecaption{
Pairs of running couplings at fixed bare coupling
$\beta=4/g_0^2$
}
\vskip1ex
$$\vbox{\settabs\+  x1.0000xxx
                 &  xx10xxx
                 &  xx1.0000(00)xxx
                 &  xx1.000(00)x
                 &  \cr
\vskip0.5ex
\thicktablerule
\vskip1ex
                \+  \hfill $\beta$ \hfill
                 &  \hfill $L/a$ \hfill
                 &  \hfill $\gbar^2(L)$ \hfill
                 &  \hfill $\gbar^2(2L)$ \hfill
                 &  \cr
\vskip 0.5ex
\thintablerule
\vskip 1ex
\+ \hfill$ 3.4564 $\hfill & \hfill$  @5        $\hfill
                          & \hfill$ 2.0371(32) $\hfill
                          & \hfill$ 2.413(15)  $\hfill & \cr
\+ \hfill$ 3.5408 $\hfill & \hfill$  @6        $\hfill
                          & \hfill$ 2.0369(52) $\hfill
                          & \hfill$ 2.418(16)  $\hfill & \cr
\+ \hfill$ 3.6045 $\hfill & \hfill$  @7        $\hfill
                          & \hfill$ 2.0370(55) $\hfill
                          & \hfill$ 2.397(19)  $\hfill & \cr
\+ \hfill$ 3.6566 $\hfill & \hfill$  @8        $\hfill
                          & \hfill$ 2.0370(63) $\hfill
                          & \hfill$ 2.447(17)  $\hfill & \cr
\+ \hfill$ 3.7425 $\hfill & \hfill$   10       $\hfill
                          & \hfill$ 2.0369(83) $\hfill
                          & \hfill$ 2.426(22)  $\hfill & \cr
\vskip 1ex
\+ \hfill$ 3.1898 $\hfill & \hfill$  @5        $\hfill
                          & \hfill$ 2.3800(43) $\hfill
                          & \hfill$ 2.981(23)  $\hfill & \cr
\+ \hfill$ 3.2751 $\hfill & \hfill$  @6        $\hfill
                          & \hfill$ 2.3801(67) $\hfill
                          & \hfill$ 2.942(21)  $\hfill & \cr
\+ \hfill$ 3.3428 $\hfill & \hfill$  @7        $\hfill
                          & \hfill$ 2.3799(67) $\hfill
                          & \hfill$ 2.968(26)  $\hfill & \cr
\+ \hfill$ 3.4009 $\hfill & \hfill$  @8        $\hfill
                          & \hfill$ 2.3801(79) $\hfill
                          & \hfill$ 2.954(23)  $\hfill & \cr
\+ \hfill$ 3.5000 $\hfill & \hfill$   10       $\hfill
                          & \hfill$ 2.380(11)@ $\hfill
                          & \hfill$ 2.870(30)  $\hfill & \cr
\vskip 1ex
\+ \hfill$ 2.9568 $\hfill & \hfill$  @5        $\hfill
                          & \hfill$ 2.8401(56) $\hfill
                          & \hfill$ 3.783(33)  $\hfill & \cr
\+ \hfill$ 3.0379 $\hfill & \hfill$  @6        $\hfill
                          & \hfill$ 2.8401(91) $\hfill
                          & \hfill$ 3.731(35)  $\hfill & \cr
\+ \hfill$ 3.0961 $\hfill & \hfill$  @7        $\hfill
                          & \hfill$ 2.840(10)@ $\hfill
                          & \hfill$ 3.709(31)  $\hfill & \cr
\+ \hfill$ 3.1564 $\hfill & \hfill$  @8        $\hfill
                          & \hfill$ 2.840(11)@ $\hfill
                          & \hfill$ 3.663(34)  $\hfill & \cr
\+ \hfill$ 3.2433 $\hfill & \hfill$   10       $\hfill
                          & \hfill$ 2.841(16)@ $\hfill
                          & \hfill$ 3.695(43)  $\hfill & \cr
\vskip 1ex
\+ \hfill$ 2.7124 $\hfill & \hfill$  @5        $\hfill
                          & \hfill$ 3.550(10)@ $\hfill
                          & \hfill$ 5.456(40)  $\hfill & \cr
\+ \hfill$ 2.7938 $\hfill & \hfill$  @6        $\hfill
                          & \hfill$ 3.550(14)@ $\hfill
                          & \hfill$ 5.287(43)  $\hfill & \cr
\+ \hfill$ 2.8598 $\hfill & \hfill$  @7        $\hfill
                          & \hfill$ 3.550(15)@ $\hfill
                          & \hfill$ 5.310(58)  $\hfill & \cr
\+ \hfill$ 2.9115 $\hfill & \hfill$  @8        $\hfill
                          & \hfill$ 3.550(16)@ $\hfill
                          & \hfill$ 5.168(38)  $\hfill & \cr
\+ \hfill$ 3.0071 $\hfill & \hfill$   10       $\hfill
                          & \hfill$ 3.550(23)@ $\hfill
                          & \hfill$ 5.122(58)  $\hfill & \cr
\vskip 1ex
\thicktablerule}$$
\endinsert

For the heatbath part of the cycle, a modified
Creutz algorithm [\ref{FabHan}] was implemented. The local
heatbath $\SUtwo$ measure cannot be generated exactly but requires
an accept/reject step.
The observed acceptance rate
increases towards smaller couplings
and was always greater than 90\%.

The parameter $\Nor$ generally
has to grow as one approaches the continuum
limit. In various models
a tuning roughly
inversely proportional to the lattice spacing proves to
be optimal [\ref{ORtuningI},\ref{ORtuningII}].
Some of the integrated autocorrelation
times
achieved in this way
are shown in fig.~1.
All points refer to approximately
the same renormalized coupling
$\gbar^2(L)\simeq3.7$, i.e.~the measurements
are made at ``constant physics".
At large $L/a$ the autocorrelation time is proportional to
$(L/a)^z$ (dashed line). The fit gives
$z=1.0(1)$ for the dynamical critical exponent, which
coincides with the expected best value for over-relaxed algorithms.

\subsection 3.3 Simulation results

As explained in subsect.~2.6, the step scaling function
$\sigma(s,u)$
may be computed by simulating pairs of
lattices at the same bare coupling with sizes
$L$ and $L'=sL$.
In our calculations we chose $s=2$ throughout
and produced enough data to be able to perform
4 renormalization steps in succession (see table 2).
We have, furthermore, done an additional set of
simulations at a fixed large value of the renormalized
coupling (table 3). These data will be used in sect.~4 to
make contact with the low-energy scales of the theory.

In each block of data listed in table 2,
the renormalized coupling on the
smaller lattices is constant within errors,
i.e.~the bare coupling has been tuned so as to achieve this.
A reweighting technique was employed in this step,
as discussed in subsect.~4.1 of
ref.[\ref{LueWeWo}].


\topinsert
\newdimen\digitwidth
\setbox0=\hbox{\rm 0}
\digitwidth=\wd0
\catcode`@=\active
\def@{\kern\digitwidth}
\tablecaption{
Bare coupling vs.~lattice size
at fixed $\gbar^2(L)=4.765$
}
\vskip1ex
$$\vbox{\settabs\+  xxx$10$
                 &  xxx$1.0000(00)$
                 &  xxxxxx$10$
                 &  xxx$1.0000(00)$
                 &  x\cr
\vskip0.5ex
\thicktablerule
\vskip1ex
                \+  \hfill $L/a$              \hskip0.0ex
                 &  \hfill $\beta$            \hskip5.0ex
                 &  \hfill $L/a$              \hskip0.0ex
                 &  \hfill $\beta$            \hskip5.0ex
                 &  \quad\cr
\vskip 0.5ex
\thintablerule
\vskip 1ex
  \+ \hfill$ 6$ & \hfill$2.5752(28)$ &
     \hfill$10$ & \hfill$2.7824(22)$ & \cr
  \+ \hfill$ 7$ & \hfill$2.6376(20)$ &
     \hfill$12$ & \hfill$2.8485(32)$ & \cr
  \+ \hfill$ 8$ & \hfill$2.6957(21)$ &
     \hfill$14$ & \hfill$2.9102(62)$ & \cr
\vskip 1ex
\thicktablerule}$$
\endinsert

The statistical errors were estimated both by jackknife binning
(using several 100 bins)
and, when no reweighting was required,
by summing the autocorrelation
function over an appropriate time interval.
In the first three
series of table~2 about
35 hours of CPU time were spent for
the lattices with $L'/a\leq14$,
60 hours for $L'/a=16$ and
140 hours for the largest system with $L'/a=20$.
All times were roughly doubled in the last series
(where $\gbar^2(L)=3.55$).
The typical speed achieved by our program is
210 Mflop/s. Speed and CPU times
refer to a single CRAY YMP processor.


\section 4. Data Analysis and Results

We now follow the strategy sketched in subsect.~2.7 and determine
the running coupling in the continuum theory
for a range of box sizes $L$ given in physical units.

\topinsert
\vbox{\vskip 12 true cm plus 0.2 true cm}
\figurecaption{
Extrapolation of the lattice step scaling function
$\Sigma(2,u,a/L)$ to the continuum limit.
The left-most points represent the extrapolated values as
given in table 4.
}
\endinsert

\subsection 4.1 Step scaling function

The simulation results listed above allow us to compute the
step scaling function $\sigma(2,u)$ at 4 values of $u$.
This involves an extrapolation of the lattice data to
the continuum limit $a/L\to0$ (see fig.~2).
The cutoff effects that we observe are quite small
and linearly decreasing with the lattice spacing.
This is the theoretically expected behaviour.
At the lower values of the coupling, the size of the
effect is compatible with what one predicts from
perturbation theory
(cf.~subsect.~2.6).
For a detailed comparision a higher statistical
precision would however be required.

To perform the extrapolation to the continuum limit,
the error on the argument $u$ has been traded for an
additional error on $\Sigma$, using an approximate value for
$\partial\Sigma/\partial u$.
A straightforward linear fit then yields the values
and errors quoted in table 4.
These can be compared with what one obtains by integrating the
renormalization group equation (2.19),
taking $u$ as the initial value and the 2--loop formula
for the Callan-Symanzik $\beta$--function (third column in table 4).
The agreement is perfect, except for a $2\,\sigma$ deviation
at the lowest value of $u$.
In a set of 4 independent measurements,
this is not an unlikely event, however.

We shall soon discover that
the couplings occurring in
table 4 correspond to a range of box sizes $L$ from about
$0.023\,\fm$ to $0.33\,\fm$.
It is thus rather surprising that
the step scaling function is so accurately reproduced by
perturbation theory.

\topinsert
%
\newdimen\digitwidth
\setbox0=\hbox{\rm 0}
\digitwidth=\wd0
\catcode`@=\active
\def@{\kern\digitwidth}
\tablecaption{
Values of the step scaling function
}
\vskip1ex
$$\vbox{\settabs\+  x1.000xxxx
                 &  xxxx1.00(00)xxx
                 &  xxxx1.00(00)xxx
                 &  \cr
\vskip0.5ex
\thicktablerule
\vskip1ex
                \+  \hfill $u$ \hfill
                 &  \hfill $\sigma(2,u)$\hfill
                 &  \hfill $\sigma(2,u)_{\rm 2-loop}$\hfill
                 &  \cr
\vskip 0.5ex
\thintablerule
\vskip 1ex
\+ \hfill$ 2.037 $\hfill & \hfill$   2.45(4)@    $\hfill
                         & \hfill$   2.38        $\hfill & \cr
\+ \hfill$ 2.380 $\hfill & \hfill$   2.84(6)@    $\hfill
                         & \hfill$   2.86        $\hfill & \cr
\+ \hfill$ 2.840 $\hfill & \hfill$   3.54(8)@    $\hfill
                         & \hfill$   3.58        $\hfill & \cr
\+ \hfill$ 3.550 $\hfill & \hfill$   4.76(12)    $\hfill
                         & \hfill$   4.83        $\hfill & \cr
\vskip 1ex
\thicktablerule}$$
\endinsert

\topinsert
%
\newdimen\digitwidth
\setbox0=\hbox{\rm 0}
\digitwidth=\wd0
\catcode`@=\active
\def@{\kern\digitwidth}
\tablecaption{
Running coupling at scales given in units of $L_8$
}
\vskip1ex
$$\vbox{\settabs\+  x1.000(00)xxx
                 &  xx1.000x
                 &  \cr
\vskip0.5ex
\thicktablerule
\vskip1ex
                \+  \hfill $L/L_8$ \hfill
                 &  \hfill $\gbar^2(L)$\hfill
                 &  \cr
\vskip 0.5ex
\thintablerule
\vskip 1ex
\+ \hfill$ 1.000@@@@ $\hfill & \hfill $4.765$ \hfill & \cr
\+ \hfill$ 0.500(23) $\hfill & \hfill $3.550$ \hfill & \cr
\+ \hfill$ 0.249(19) $\hfill & \hfill $2.840$ \hfill & \cr
\+ \hfill$ 0.124(13) $\hfill & \hfill $2.380$ \hfill & \cr
\+ \hfill$ 0.070(8)@ $\hfill & \hfill $2.037$ \hfill & \cr
\vskip 1ex
\thicktablerule}$$
\endinsert

\topinsert
\newdimen\digitwidth
\setbox0=\hbox{\rm 0}
\digitwidth=\wd0
\catcode`@=\active
\def@{\kern\digitwidth}
\tablecaption{
Values of the string tension and the box size $L_8$
}
\vskip1ex
$$\vbox{\settabs\+  xxx$2.85$
                 &  xxxxx$0.00354(26)$
                 &  xxxxxx$12.00(0)$
                 &  xxxxx$0.713(4)$
                 &  \cr
\vskip0.5ex
\thicktablerule
\vskip1ex
                \+  \hfill $\beta$               \hskip3.0ex
                 &  \hfill $a^2\stringtension$   \hskip6.5ex
                 &  \hfill $L_8/a$               \hskip5.0ex
                 &  \hfill $L_8\sqrt{K}$         \hskip4.0ex
                 &  \cr
\vskip 0.5ex
\thintablerule
\vskip 1ex
  \+ \hfill $2.70           $ \hfill &
     \hfill $0.0103(2)@@    $ \hfill &
     \hfill $@8.08(4)       $ \hfill &
     \hfill $0.820(9)@      $ \hfill & \cr
  \+ \hfill $2.85           $ \hfill &
     \hfill $0.00354(26)    $ \hfill &
     \hfill $11.98(7)       $ \hfill &
     \hfill $0.713(26)      $ \hfill & \cr
\vskip 1ex
\thicktablerule}$$
\endinsert

\subsection 4.2 Running coupling

For the error analysis it proves useful
to set up the recursion (2.26)
in a logically reversed manner, where one first
specifies the sequence of couplings
$u_i$, $i=0,1,\ldots$, and then computes the associated
scale factors $s_i$.
The couplings are defined by
$$
  \eqalign{
  u_0&=2.037,
  \quad\qquad
  u_1=\sigma(2,u_0), \cr
  \noalign{\vskip1ex}
  u_2&=2.380,
  \quad\qquad
  u_3=\sigma(2,u_2), \cr
  \noalign{\vskip1ex}
  u_4&=2.840,
  \quad\qquad
  u_5=\sigma(2,u_4), \cr
  \noalign{\vskip1ex}
  u_6&=3.550,
  \quad\qquad
  u_7=\sigma(2,u_6), \cr
  \noalign{\vskip1ex}
  u_8&=4.765.}
  \eqno\enum
$$
$u_0,u_2,u_4$ and $u_6$ coincide with the numbers
listed in the first column of table 4.
To the precision stated, $u_1,u_3,u_5$ and $u_7$
are thus given by the second column in this table.


By definition, the scale factors $s_i$
[as determined through eq.(2.26)]
are exactly equal to $2$
for $i=0,2,4$ and $6$.
In all other cases, $s_i$ is a number close to 1, which may be
computed by
evolving the coupling from $u_i$ to $u_{i+1}$ using
the 2--loop formula for the
$\beta$--function.
Since the step scaling function is well reproduced by
perturbation theory, we estimate that
the systematic errors incurring at this point
are negligible compared to the statistical errors in table 4
(which translate to
errors on $s_1,s_3,s_5$ and $s_7$).


The box sizes
$L_i$ at which $\gbar^2(L_i)=u_i$ can now be computed
straightforwardly through eq.(2.27).
The result is given in table 5 and will be discussed below,
after converting to more physical units.
In this computation the errors on the scale factors $s_i$
have been added in quadrature,
because they arise from independent simulations.

\topinsert
\vbox{\vskip 10 true cm plus 0.2 true cm}
\figurecaption{
Bare coupling $\beta=4/g_0^2$
versus lattice size at fixed
$\gbar^2(L)=4.765$.
The dashed curve is a fit,
$\beta=1.905+0.380\times\ln(L/a)$, to the data points with
$L/a\geq8$.
}
\endinsert

\topinsert
\vbox{\vskip 12 true cm plus 0.2 true cm}
\figurecaption{
Comparision of numerically computed values of the running coupling
(data points) with perturbation theory.
The dashed (dotted) curve is obtained by
integrating the evolution equation (2.19), starting
at the right-most point and using the 2--loop (1--loop) formula
for the $\beta$--function.
}
\endinsert

\subsection 4.3 Physical scales and computation of $\alpha_{\msbar}(q)$

At the largest coupling in the recursion (4.1),
it is possible to
make contact with the low-energy scales of the theory in
infinite volume.
As already mentioned in sect.~2, we decided to take
the string tension $\stringtension$ as a reference
energy scale in this regime.
$\stringtension$ has been determined on large lattices
at $\beta=2.70$ [\ref{MiPe}] and more recently at
$\beta=2.85$ [\ref{UKQCDI}] (second column of table 6;
the errors quoted there are statistical only)
\footnote{$\dag$}{\footnotefont
The numbers published in
refs.[\ref{MiPe},\ref{UKQCDI}]
are actually higher by a factor of $1.09$ and
$1.24$, respectively.
The string tension has meanwhile
been reevaluated, the results being as given here.
We thank C.~Michael and the UKQCD collaboration
for communicating these revisions to us.}.
%


To determine the dimensionless combination
$L_8\sqrt{K}$, we also need $L_8$ in lattice units at the
same values of the bare coupling.
This information can be extracted easily
by interpolating
the data listed in
table 3 (see fig.~3).
The outcome of the calculation
is given in the third and fourth columns of table 6.

The difference between the values of $L_8\sqrt{K}$ obtained
at $\beta=2.70$ and $\beta=2.85$ can tentatively be interpreted
as a cutoff effect.
In principle, $L_8\sqrt{K}$ should be extrapolated to the
continuum limit in the same way as the
step scaling function.
We, however, are hesitating to do this, because we have
only two data points and since it is not certain that
the observed variation of $L_8\sqrt{K}$
is a pure cutoff effect.
In particular, systematic errors
on the string tension values quoted in table 6
of the order of $5-10\%$
cannot be excluded at present
[\ref{MiLetter}].

In the following we take
$L_8\sqrt{K}=0.713$ and
keep in mind that
the total error on this number could be as large as $10\%$.
If we set $\sqrt{K}=425\,\MeV$ to convert to more physical units,
we then deduce that $L_8=0.33\,\fm$.
The lower end of the range of box sizes covered by table 5
is hence roughly equal to $0.023\,\fm$.

As shown in fig.~4 the evolution of the running coupling
$\alpha(q)$ is well described by perturbation theory,
down to very low energies. The error bars in this plot
only represent the statistical errors as given in table 5,
but not the overall scale uncertainty discussed above.
It should be emphasized
that the latter amounts to a multiplication of
the energy scale by a {\it constant} factor
and so has no bearing on the scaling properties
of the coupling.

At the highest energies reached,
we can finally convert to the $\msbar$
scheme of dimensional regularization
using perturbation theory.
A typical result is
$$
  \alpha_{\msbar}(q)=
  0.187\pm0.005\pm0.009
  \quad\hbox{at}\quad
  q=20\times\sqrt{K}.
  \eqno\enum
$$
The first error here is statistical, as inferred from table 5,
while the second is an estimate of the total systematic error
arising from
a possible order $\alpha^3$ correction in eq.(2.15)
and the $10\%$ scale uncertainty mentioned above.

\subsection 4.4 Relation between the bare
                and the renormalized coupling

Let us consider a large lattice,
with bare coupling $g_0^2$
deep in the scaling region, and let us assume that
the lattice spacing $a$ is known in units of some low-energy
scale.
In the {\it continuum} theory,
the running coupling $\gbar^2(a)$
then is a well-determined quantity, which may be
related to the lattice coupling through an asymptotic series,
$$
  \gbar^2(a)=g_0^2+c_1g_0^4+c_2g_0^6+\ldots,
  \eqno\enum
$$
with purely numerical coefficients.
At present the 1--loop coefficient is known
[\ref{LueNaWeWo}],
$$
  c_1=0.20235,
  \eqno\enum
$$
and an effort is being made to extend the calculation to
the next order
[\ref{TwoLoop}].
Whether the expansion applies in the accessible range of
bare couplings
(at $\beta=2.85$ for example)
is not known, however,
and one may in fact have serious doubts that it does,
because the first order correction is
uncomfortably large.

Other renormalized
couplings fare no better in this respect
ond one is thus led to suspect that $g_0^2$
is a ``bad" expansion parameter
[\ref{Pa}--\ref{LeMaII}].
On the basis of a mean field argument,
Parisi [\ref{Pa}]
suggested many years ago that
$$
  \gtilde^2=g_0^2/P,
  \qquad
  P=\frac{1}{2}\left\langle\tr\,U(p)\right\rangle,
  \eqno\enum
$$
would be
a more natural choice of bare parameter for the lattice theory.
The plaquette expectation value $P$ is
to be computed on an infinite lattice.
At $\beta=2.70$ and $\beta=2.85$,
it is equal to 0.68558 and 0.70577, respectively
[\ref{CaCuGiPa}].
In terms of $\gtilde^2$, eq.(4.3) becomes
$$
  \gbar^2(a)=\gtilde^2+\tilde{c}_1\gtilde^4
                      +\tilde{c}_2\gtilde^6+\ldots,
  \eqno\enum
$$
with
$$
  \tilde{c}_1=c_1-\frac{3}{16}=0.01485.
  \eqno\enum
$$
The corresponding
expansion of the renormalized coupling
in the $\msbar$ scheme of dimensional regularization
plays a key r\^ole in the work
of El-Khadra et al.[\ref{AidaEtAl}] and is further discussed
in ref.[\ref{LeMaII}].

While the 1--loop coefficient $\tilde{c}_1$ is much smaller
than $c_1$, it is not guaranteed that
the higher order corrections are small
and so it remains unclear whether
the series (4.6) yields a reliable estimate for the
renormalized coupling.
Using the results obtained above we are now in a position
to answer this question.
At $\beta=2.85$, for example,
the lattice spacing in units of $L_8$ is equal to $0.0835(5)$,
which is in the range covered by table 5.
For the running coupling, the value
$\gbar^2(a)=2.11(5)$ is thus obtained.
This is to be compared with the rhs of eq.(4.6), which
evaluates to $\gbar^2(a)=2.05$ at 1--loop order.
So we do confirm that
the higher order corrections
are small and thus conclude that $\gtilde^2$ is a good expansion
parameter at the scale of the cutoff.


\section 5. Conclusions

Lattice gauge theories have been invented to study
the properties of Yang-Mills theories
and QCD at low energies.
It proved to be difficult, however,
to make contact with
the perturbative regime, where
weakly interacting quarks and gluons are the important degrees
of freedom.
The obstacle is that one cannot easily hold a wide range
of physical scales on a single lattice, at least as long
as numerical simulations are the only practical way
to do non-perturbative computations in the scaling region.

Using a recursive finite-size technique, we have now been
able to close this gap in the case of the pure $\SUtwo$
gauge theory.
We found
that the evolution of the renormalized coupling
in the chosen scheme
is well described by perturbation theory
over the whole range of energies covered (cf.~fig.~4).
This is a bit surprising, but
it should be noted that
the coupling is defined through an off-shell amplitude
and so is insensitive to threshold effects.
In any case, our result proves that there is no
complicated ``intermediate" energy range before
the coupling becomes small and slowly decreasing according
to the perturbative renormalization group.
In this respect the situation is as in the
two-dimensional non-linear $\sigma$--model
[\ref{LueWeWo}].

Our method allows us to compute the running coupling
in say the $\msbar$ scheme of dimensional regularization
at energies far above the masses of the light particles in the
theory. An example of such a result is given in eq.(4.2).
It is certainly possible to achieve a higher precision
in this calculation. Compared to the power of present day
parallel computers, we have used an only small
amount of CPU time. A refined study, with more statistics
and an $\rmO(a)$ improved action
[\ref{LueNaWeWo}],
is hence clearly feasible.
It is then also necessary
to extend the series (2.15) to the 2--loop level
[\ref{TwoLoop}]
to keep the balance between systematic and statistical errors.

We finally note that an extension of our work to the
$\SUthree$ Yang-Mills theory should not meet any fundamental
difficulty.
QCD requires more thought, however,
because
the scale dependence
of the quark masses must be taken into account.

We are indebted to F.~Gutbrod for discussions
and A.~Kronfeld and C.~Michael for correspondence.
The computations have been performed on the CRAY computers at
HLRZ and CERN. We thank these institutions for their support.



\beginbibliography

\bibitem{Wilson}
K. G. Wilson,
Monte Carlo calculations for the lattice gauge theory,
{\it in}\/: Recent developments in gauge theories
(Carg\`ese 1979),
ed. G. 't Hooft et al. (Plenum, New York, 1980)

\bibitem{LueWeWo}
M. L\"uscher, P. Weisz and U. Wolff,
Nucl. Phys. B359 (1991) 221

\bibitem{LueNaWeWo}
M. L\"uscher, R. Narayanan, P. Weisz and U. Wolff,
The Schr\"odinger functional --- a renormalizable probe for
non-Abelian gauge theories,
DESY preprint 92-025 (1992), to appear in Nucl. Phys. B

\bibitem{AidaEtAl}
A. E. El-Khadra, G. Hockney, A. S. Kronfeld and P. B. Mackenzie,
A determination of the strong coupling constant from
the charmonium spectrum,
Fermilab preprint 91/354-T (1992)

\bibitem{MiGbar}
C. Michael,
The running coupling from lattice gauge theory,
Liverpool preprint LTH 279 (1992)

\bibitem{BaBuDuMu}
W. A. Bardeen, A. Buras, D. W. Duke and T. Muta,
Phys. Rev. D18 (1978) 3998

\bibitem{Fi}
W. Fischler,
Nucl. Phys. B129 (1977) 157

\bibitem{Bi}
A. Billoire,
Phys. Lett. B92 (1979) 343

\bibitem{GroWi}
D. J. Gross and F. A. Wilczek,
Phys. Rev. Lett. 30 (1973) 1343

\bibitem{Pol}
H. D. Politzer,
Phys. Rev. Lett. 30 (1973) 1346

\bibitem{Jo}
D. R. T. Jones,
Nucl. Phys. B75 (1974) 531

\bibitem{Cas}
W. E. Caswell, Phys. Rev. Lett. 33 (1974) 244

\bibitem{hit}
G. Parisi, R. Petronzio and F. Rapuano,
Phys. Lett. B128 (1983) 418

\bibitem{ORI}
M. Creutz,
Phys. Rev. D36 (1987) 515

\bibitem{ORII}
F. R. Brown and T. J. Woch,
Phys. Rev. Lett. 58 (1987) 2394

\bibitem{ORIII}
K. Decker and Ph. de Forcrand,
Nucl. Phys. B (Proc. Suppl.) 17 (1990) 567

\bibitem{FabHan}
K. Fabricius and O. Haan, Phys. Lett. B143 (1984) 459

\bibitem{ORtuningI}
J. Apostolakis, G. C. Fox and C. F. Baillie,
Nucl. Phys. B (Proc. Suppl.) 20 (1991) 678

\bibitem{ORtuningII}
U. Wolff,
Dynamics of hybrid over-relaxation in the
Gaussian model,
preprint CERN-TH 6408/92, to appear in Phys. Lett. B

\bibitem{MiPe}
C. Michael and S. Perantonis,
Nucl. Phys. (Proc. Suppl.) 20 (1991) 177

\bibitem{UKQCDI}
S. P. Booth et al. (UKQCD Collab.),
Phys. Lett. B275 (1992) 424

\bibitem{MiLetter}
C. Michael,
private communication

\bibitem{Pa}
G. Parisi,
{\it in}\/: High-Energy Physics --- 1980,
XX. Int. Conf., Madison (1980), ed. L. Durand and L. G. Pondrom
(American Institute of Physics, New York, 1981)

\bibitem{Hasenfratz}
A. Hasenfratz and P. Hasenfratz,
Phys. Lett. 93B (1980) 165; Nucl. Phys. B193 (1981) 210

\bibitem{LeMaI}
G. P. Lepage and P. B. Mackenzie,
Nucl. Phys. B (Proc. Suppl.) 20 (1991) 173

\bibitem{LeMaII}
G. P. Lepage and P. B. Mackenzie,
Fermilab preprint 91/355-T (1992), to appear

\bibitem{CaCuGiPa}
M. Campostrini, G. Curci, A. Di Giacomo and G. Paffuti,
Z. Phys. C32 (1986) 377

\bibitem{TwoLoop}
M. L\"uscher, R. Narayanan, A. van de Ven, C. Vohwinkel, P. Weisz
and U. Wolff,
work in progress

\endbibliography

\bye